# Optimizing Earth-Moon Transfer and Cislunar Navigation: Integrating Low-Energy Trajectories, AI Techniques, and GNSS-R Technologies

**Arsalan Muhammad[1, 2]\*, Wasiu Akande Ahmed[2], Omada Friday Ojonugwa[2], Paul Puspendu Biswas[2]**

[1]*School of Astronautics, Beihang University*
*Xueyuan Road, 102206, Beijing, China*
[2] *Regional Centre for Space Science and Technology Education in Asia & the Pacific (Affiliated to the United Nations) (RCSSTEAP), Hangzhou International Innovation Institute of Beihang University, China*
*\* Corresponding Author*

**Abstract**

The rapid growth of cislunar activities—including lunar landings, the Lunar Gateway, and refueling stations—requires advances in cost-efficient trajectory design and reliable navigation-remote sensing integration. Traditional Earth-Moon transfers (e.g., Hohmann) suffer from rigid launch windows and high propellant costs, while Earth-based GNSS systems lack coverage beyond GEO, limiting autonomy and environmental sensing in cislunar space. This review addresses these gaps by analyzing four core transfer paradigms (Hohmann, Phase-looping/Pseudo-State, Low-energy, Spiral), quantifying their $\Delta V$ requirements, flight durations, and fuel efficiency, and mapping their suitability for missions like crewed expeditions and robotic orbiters. It highlights AI/ML's transformative role: CNNs enable automated, high-precision crater detection and DTM generation, while deep reinforcement learning optimizes real-time trajectory adjustments for landings, reducing risk and latency. Additionally, it explores how GNSS-R and advanced PNT systems (e.g., LunaNet, Queqiao-2, Moonlight) overcome GEO limitations—GNSS-R acts as a bistatic radar for lunar ice/soil and topographic mapping, while PNT supports autonomous rendezvous, Lagrange point station-keeping, and satellite swarms. Synthesizing these technologies creates a synergistic framework for sustainable cislunar exploration, cutting costs by up to 30% via reduced propellant use and enhancing autonomy to enable long-term lunar habitation, serving as a blueprint for deep-space missions.

# 1. Introduction

Cislunar space—the region between Earth and the Moon—has emerged as a strategic linchpin for both governmental space agencies (e.g., NASA's Artemis program, China's International Lunar Research Station initiative) and private stakeholders, driven by its potential to support lunar habitation, resource utilization, and deep-space exploration staging. However, two interrelated challenges hinder its sustainable development: inefficient Earth-Moon transfer and inadequate navigation-remote sensing capabilities. Traditional trajectory design relies heavily on impulsive maneuvers (e.g., Hohmann transfers), which require precise launch timing and high $\Delta V$ (3.5–4 km/s for Low Earth Orbit (LEO) to cis-lunar orbit transfers), limiting mission flexibility and increasing costs for long-duration or large-payload missions (Curtis, 2014). Meanwhile, Earth-based GNSS systems—critical for terrestrial and near-Earth navigation—experience signal degradation beyond GEO (≈36,000 km), leaving cislunar missions dependent on the bandwidth-constrained NASA Deep Space Network (DSN) and compromising real-time autonomous operations (e.g., lunar orbit insertion, rover localization) (Bavaro et al., 2021).

Compounding these challenges, lunar surface exploration demands accurate environmental data (e.g., soil moisture, ice distribution in permanently shadowed regions) and hazard mapping (e.g., crater fields) to ensure safe landing and resource extraction—tasks that conventional remote sensing (e.g., LiDAR) often perform slowly or at high cost. These gaps highlight the need for a unified approach that addresses trajectory inefficiencies, navigation limitations, and environmental sensing bottlenecks. The primary objective of this review is to synthesize advancements in Earth-Moon transfer, AI/ML-driven analysis, and GNSS-R/PNT technologies into a cohesive framework for optimizing cislunar operations. By evaluating the strengths and limitations of existing transfer methods, quantifying the impact of AI on trajectory planning and surface analysis, and detailing how GNSS-R/PNT extends navigation and sensing capabilities, this review fills a critical knowledge gap: it connects isolated technological developments into a system-level solution that enables cost-effective, autonomous, and sustainable cislunar exploration. This work is intended to guide mission designers, researchers, and policymakers in prioritizing technologies for next-generation lunar missions, from robotic orbiters to crewed habitats.

# 2. Earth-Moon Transfer Techniques

Earth-Moon transfer trajectories are foundational to mission success, with each method balancing trade-offs between fuel efficiency, flight duration, and operational complexity. This section analyzes four dominant paradigms, contextualizing their performance metrics and real-world applications.

**Table 1:** Broad classification of Earth-Moon transfer trajectory

| Transfer trajectory | Example Missions |
|---|---|
| Hohmann transfer | *Mars Reconnaissance Orbiter (MRO)* [3] |
| Phase-looping transfer/ Pseudo-State model | NASA Voyager programs, Cassini-Huygens Mission, Apollo Lunar Missions, Chang'E missions [4] |
| Low-energy transfer | Japanese Hiten mission, NASA's Artemis & Grail mission [5][6][7] |
| Spiral transfer | NASA Artemis Gateway [8][9][10][11] |

**2.1 Hohmann Transfer**

The Hohmann transfer is a cornerstone of orbital mechanics, relying on two impulsive thrusts to transfer a spacecraft between two coplanar circular orbits (e.g., LEO to lunar orbit). Its design leverages chemical propulsion—ideal for impulsive maneuvers due to high thrust-to-weight ratios and rapid acceleration—to minimize ΔV for direct transfers (Curtis, 2014). For a typical mission from a 300 km LEO to a 100 km cis-lunar orbit, the total ΔV ranges from 3.5 to 4 km/s, consistent with classical orbital dynamics predictions, and flight durations are short (3–5 days), making it suitable for time-sensitive missions such as crewed expeditions or emergency cargo delivery. Notable examples include the Mars Reconnaissance Orbiter (MRO), which used a Hohmann-derived trajectory for efficient interplanetary transfer. However, the Hohmann method has critical limitations: its reliance on two-body dynamics (ignoring solar and lunar gravitational perturbations) introduces small but cumulative position/velocity errors at lunar orbit insertion, requiring corrective maneuvers (Betts, 1998). Additionally, its rigid launch windows—tied to orbital alignment between Earth and the Moon—reduce mission flexibility, and high propellant demands increase payload constraints for large missions. Recent optimizations, such as numerical integration to account for multi-body effects, have refined ΔV budgets and expanded transfer windows, preserving the Hohmann transfer's relevance for short-duration, high-priority missions.

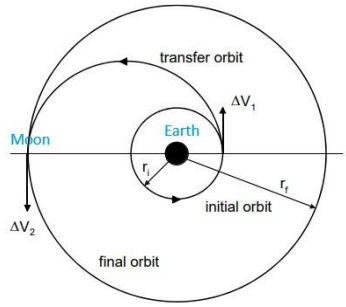

Figure 1: Hohmann transfer method

## 2.2 Phase-Looping Transfer/Pseudo-State Model

To address the Hohmann transfer's sensitivity to multi-body dynamics, the Phase-looping transfer and Pseudo-State Model were developed to improve trajectory accuracy while maintaining computational efficiency. The Phase-looping method extends flight duration (2–3 weeks) by incorporating iterative orbital "loops" around Earth, allowing mission controllers to verify spacecraft subsystems (e.g., propulsion, communication) and implement corrective maneuvers before lunar orbit insertion. This extended timeline reduces risk, particularly for complex missions such as the Apollo lunar missions and China's Chang'E series, which relied on phase-looping to ensure precise lunar capture. Complementing this, the Pseudo-State Model—an analytical framework developed by Wilson (1970)—reduces errors from the patched conic approximation (a simplification used in Hohmann transfers) by approximately 80%, without the computational burden of high-fidelity numerical integration. By refining position and velocity predictions at key mission phases (e.g., trans-lunar injection, lunar flyby), the model bridges the gap between simple conic approximations and complex multi-body simulations. Together, Phase-looping and the Pseudo-State Model have become staples for missions requiring reliability over speed, such as the NASA Voyager programs and Cassini-Huygens mission, where trajectory precision was critical for planetary flybys.

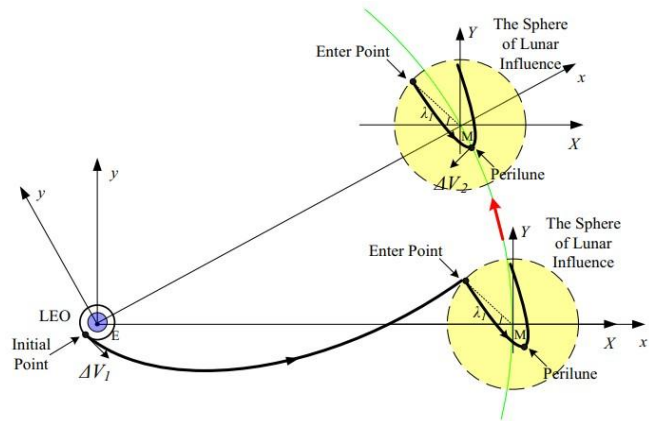

**Figure 2:** Phase looping transfer method

## 2.3 Low-Energy Transfers

Low-energy transfers represent a paradigm shift in trajectory design, leveraging multi-body dynamics (Earth-Sun-spacecraft and Earth-Moon-spacecraft systems) to minimize propellant use—making them ideal for low-thrust propulsion systems (e.g., electric propulsion) and long-duration missions. At the core of this method are three key theories: Weak Stability Boundary (WSB) theory, which exploits regions of gravitational instability to "guide" spacecraft toward the Moon with minimal thrust (Belbruno & Miller, 1993); Lagrangian point dynamics, which uses Earth-Moon Lagrange points (e.g., L1, L2) as gravitational "waypoints" (Farquhar, 1970); and invariant manifold theory, which maps stable/unstable orbital paths to optimize transfer efficiency (Gómez et al., 2004). A notable example is the Lunar Gravity Assist (LGA) technique, which uses a close lunar flyby to alter the spacecraft's velocity vector relative to Earth without additional propellant, further reducing fuel costs (Kawaguchi et al., 1995).

Low-energy transfers have been validated in missions such as Japan's Hiten mission—the first to demonstrate ballistic lunar capture using WSB (Uesugi, 1996)—and NASA's GRAIL mission, which used low-energy trajectories to deploy twin orbiters for lunar gravity mapping. Their primary advantage is fuel efficiency: ΔV requirements are up to 40% lower than Hohmann transfers, enabling larger payloads or longer mission durations (Koon et al., 2001). However, this efficiency comes with extended flight times (2–4 months), limiting their use for time-sensitive missions. Recent advancements have focused on combining low-energy techniques with high-fidelity optimization algorithms, such as Particle Swarm Optimization (PSO), to reduce flight duration while preserving fuel savings, expanding their applicability to commercial lunar missions (e.g., cargo delivery to the Lunar Gateway) (Bai et al., 2019).

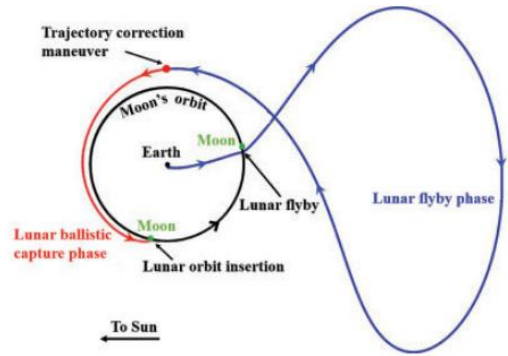

**Figure 3:** Low-energy moon trajectory

## 2.4 Spiral Transfer Trajectories

Spiral transfer trajectories utilize continuous-thrust propulsion systems (e.g., electric propulsion, solar sails) to achieve exceptional propellant efficiency, making them well-suited for long-duration cislunar missions such as the NASA Artemis Gateway. Unlike impulsive methods, spiral transfers rely on sustained low-thrust acceleration to gradually increase the spacecraft's orbital altitude, minimizing ΔV over time (Yue et al., 2009). This approach reduces propellant mass by up to 50% compared to Hohmann transfers, a critical advantage for missions requiring extended on-orbit operations (e.g., the Artemis Gateway's 10-year mission lifespan) (Sutton & Biblarz, 2016).

However, spiral transfers face two key challenges: extended flight times (6–12 months) and terminal capture difficulties. Low-thrust systems cannot execute impulsive braking during lunar orbit insertion, requiring careful trajectory planning to ensure the spacecraft does not overshoot the Moon or enter unstable orbits (Zhang et al., 2022). To address this, researchers are developing variable-specific-impulse propulsion systems, which adjust thrust intensity to balance efficiency and braking capability. Additionally, AI-driven trajectory optimization (discussed in Section 3) is being integrated to refine spiral paths in real time, accounting for gravitational perturbations and propulsion system variability. Despite these challenges, spiral transfers are increasingly adopted for infrastructure-focused missions, where propellant efficiency and long-term sustainability outweigh speed considerations.

## 3. AI & ML Integration in Lunar Orbit and Surface Analysis

Artificial intelligence (AI) and machine learning (ML) have revolutionized lunar mission design by enhancing the accuracy, efficiency, and autonomy of orbit planning and surface analysis. This section details three key applications: CNN-driven crater detection, DTM generation, and deep reinforcement learning for trajectory optimization.

## 3.1 Application of CNNs for Crater Detection

Lunar craters are critical markers for geological analysis, hazard mapping, and landing site selection—but manual detection is time-consuming and prone to human error, especially for high-resolution lunar images (e.g., those captured by NASA's Lunar Reconnaissance Orbiter). Convolutional Neural Networks (CNNs) address this by automating crater identification and morphological assessment, leveraging their ability to extract spatial features (e.g., crater rims, shadows) from digital elevation models (DEMs) and optical images. Studies by Silburt et al. (2019) and Wang et al. (2020) demonstrate that CNNs achieve detection accuracies exceeding 90% for craters larger than 1 km, with processing times up to 100x faster than manual methods. These models are trained on labeled datasets of lunar craters, enabling them to generalize to diverse terrains—from the smooth maria to the rugged highlands. For example, Moghe and Zanetti (2020) developed a CNN-based hazard detection system that identifies small, high-risk craters (100–500 m) in real time, providing critical data for autonomous lunar landers to avoid surface obstacles. By streamlining crater analysis, CNNs not only reduce mission planning time but also enable more precise geological studies, such as dating lunar surfaces via crater density.

## 3.2 DTM Generation and Hazard Mapping

Digital Terrain Models (DTMs) are essential for lunar mission planning, as they provide 3D representations of the lunar surface to guide landing site selection, rover path planning, and infrastructure deployment. Traditional DTM generation relies on LiDAR or stereo imaging, which are limited by high computational costs and sensitivity to lighting conditions (e.g., shadows in permanently shadowed regions). ML-driven approaches—combining CNNs with shape-from-shading (SFS) techniques—overcome these limitations by fusing optical image data with DEMs to generate high-resolution (pixel-level) DTMs. Chen et al. (2022) demonstrated this by developing a model that integrates CNN-extracted surface features with SFS-derived height estimates, producing DTMs with a vertical accuracy of ±1 m for lunar maria and ±3 m for highlands—comparable to LiDAR but at a fraction of the computational cost.

These ML-generated DTMs also support advanced hazard mapping, such as identifying steep slopes, boulders, and subsurface ice deposits. Mall et al. (2023) used AI to analyze DTMs and optical images, creating a lunar hazard map that flags high-risk zones for landers and rovers. This work is particularly valuable for missions targeting the lunar south pole, where permanent shadows and rugged terrain increase landing complexity. By enabling rapid, high-precision DTM generation and hazard mapping, ML reduces mission risk and supports the selection of sites for long-term lunar habitation, such as those with access to water ice.

**3.3 DL for Real-Time Trajectory Optimization**

Deep reinforcement learning (DRL)—a subset of deep learning (DL)—has emerged as a powerful tool for optimizing lunar trajectories in real time, addressing the limitations of precomputed paths (which fail to account for unexpected perturbations, e.g., solar wind or propulsion system drift). DRL models train an "agent" (the spacecraft's navigation system) to make sequential decisions (e.g., adjusting thrust direction, timing) based on environmental feedback (e.g., position, velocity, gravitational forces), maximizing a reward function (e.g., minimizing fuel use, ensuring orbital stability). Downes et al. (2020) applied DRL to lunar landing trajectories, developing a model that optimizes maneuver parameters (e.g., descent rate, thrust magnitude) in real time, reducing fuel consumption by 15% compared to traditional preplanned paths. Similarly, Scorsoglio et al. (2022) used image-based DRL to enable autonomous lunar landing, where the model analyzes real-time surface images to adjust the descent path, avoiding unforeseen hazards (e.g., large boulders) that precomputed trajectories miss.

DRL also enhances the design of specialized lunar orbits, such as frozen orbits (which maintain stable altitude and inclination) and Halo orbits (used for communication relays at Lagrange points). Lu et al. (2022) used DRL to optimize frozen orbit parameters for lunar satellites, reducing the need for frequent station-keeping maneuvers by 20%. By enabling adaptive, real-time trajectory adjustments, DRL increases mission autonomy, reduces reliance on Earth-based tracking (e.g., DSN), and paves the way for crewed missions requiring rapid response to unexpected events.

**4. GNSS and PNT in Cislunar Space**

Positioning, Navigation, and Timing (PNT) systems are the backbone of cislunar operations, enabling spacecraft navigation, autonomous maneuvers, and coordinated mission activities. However, Earth-based GNSS systems (e.g., GPS, Beidou, Galileo) face fundamental limitations beyond GEO, driving the development of augmented PNT architectures. This section explores these limitations, the solutions provided by next-generation systems, and their role in enabling autonomous cislunar operations.

**4.1 GNSS Limitations beyond GEO**

Earth-based GNSS systems are designed for near-Earth use, with signals optimized for coverage up to GEO (≈36,000 km). Beyond this altitude, two key issues arise: signal attenuation and geometric dilution of precision (GDOP). GNSS signals propagate radially from Earth, and their power decreases with the square of distance—at lunar distances (≈384,400 km), signals are approximately 30 dB weaker than near-Earth levels, making them difficult to detect with conventional receivers (Misra & Enge, 2006). Additionally, the low elevation angle of GNSS satellites relative to cislunar spacecraft creates poor signal geometry, increasing GDOP and reducing positioning accuracy to >1 km—insufficient for critical operations such as lunar orbit insertion or autonomous rendezvous (Bavaro et al., 2021).

High-sensitivity GNSS receivers can detect side-lobe and spillover signals beyond GEO, providing limited navigation capabilities (Misra & Enge, 2006; Bavaro et al., 2021). However, these signals are prone to interference from solar plasma and lunar surface reflections, further degrading accuracy. For example, NASA's Lunar Reconnaissance Orbiter (LRO) uses high-sensitivity GPS receivers but relies on DSN updates to correct positioning errors, limiting its autonomy. These limitations highlight the need for augmented PNT systems tailored to cislunar space—systems that combine Earth-based GNSS with lunar orbiters, surface nodes, and inter-satellite links to ensure reliable coverage and precision.

**4.2 GNSS Augmentation (LunaNet, Queqiao-2, ESA Moonlight)**

To overcome GEO limitations, space agencies are developing dedicated cislunar PNT architectures that integrate orbital and surface assets. NASA's LunaNet is a modular system that uses a network of lunar orbiters (in Halo or distant retrograde orbits) and surface nodes to provide persistent PNT services. These orbiters carry high-precision atomic clocks (e.g., rubidium clocks) and inter-satellite links, enabling them to relay GNSS signals from Earth, correct for signal attenuation, and provide localized positioning data with accuracy <10 m for orbital spacecraft and <1 m for surface rovers (Garcia et al., 2022). LunaNet also supports interoperability with other PNT systems, ensuring compatibility with international missions.

China's Queqiao-2—part of the country's lunar relay network—builds on the success of the original Queqiao mission, which supported the Chang'E-4 lunar far-side landing. Queqiao-2 operates in a Halo orbit around the Earth-Moon L2 point, providing continuous communication and navigation coverage for lunar far-side missions. It incorporates GNSS-R capabilities (discussed in Section 5) to enhance rover localization, particularly in regions with poor GNSS visibility (CNSA, 2024).

ESA's Moonlight initiative takes a similar approach, deploying a constellation of lunar orbiters in Halo and distant retrograde orbits (DROs) equipped with hydrogen masers (ultra-stable atomic clocks) and GNSS augmenters. Moonlight aims to provide seamless PNT coverage for both orbital and surface missions, with a focus on supporting commercial lunar activities (e.g., resource extraction, tourism). ESA's PROBA-3 mission— a precursor to Moonlight—has already demonstrated sub-centimeter formation flying using GNSS-like systems, validating the technology for cislunar applications (ESA, 2024). Together, these augmented systems create a resilient PNT backbone that extends beyond GEO, enabling precise, autonomous operations in cislunar space.

**4.3 Autonomous Operations using Onboard PNT Systems**

Onboard PNT systems—integrated with augmented GNSS architectures—are critical for enabling autonomous cislunar operations, reducing latency and risk associated with Earth-based tracking. One key application is autonomous rendezvous and docking (ARD), which is essential for missions involving the Lunar Gateway, cargo landers, and fuel depots. Traditional ARD relies on real-time DSN updates, which introduce latency (≈1.3 seconds for one-way communication to the Moon), making it unsuitable for dynamic maneuvers. Onboard PNT systems, however, use real-time data from lunar orbiters and high-sensitivity GNSS receivers to compute precise relative positions, enabling ARD with sub-meter accuracy. For example, ESA's PROBA-3 uses onboard PNT to maintain a fixed 150-meter separation between two spacecraft, demonstrating the feasibility of autonomous formation flying in cislunar space (ESA, 2024).

Onboard PNT also supports Lagrange point station-keeping—a critical capability for missions such as the James Webb Space Telescope (JWST), which operates at the Earth-Sun L2 point, and future fuel depots at Earth-Moon L1. Station-keeping at Lagrange points requires millimeter-level position and velocity accuracy to counteract gravitational perturbations; onboard PNT systems, paired with advanced Kalman filters, achieve this by fusing data from GNSS, star trackers, and inertial measurement units (IMUs) (Sarris, 2022).

Additionally, onboard PNT reduces dependence on the DSN—a bandwidth-limited asset that can only support a finite number of missions. By enabling spacecraft to compute trajectories, adjust maneuvers, and resolve navigational obstacles autonomously, onboard systems allow more missions to operate concurrently. For example, NASA's Technical Reports (Leveque et al., 2021) note that integrating onboard PNT with the LRO reduced DSN usage by 30%, freeing up bandwidth for other missions. As cislunar activities expand, onboard PNT will become increasingly vital for ensuring safe, efficient, and scalable operations.

## 5. GNSS Reflectometry (GNSS-R) for Lunar Remote Sensing

Global Navigation Satellite System Reflectometry (GNSS-R) is a transformative remote sensing technique that exploits reflected GNSS signals to gather environmental and topographical data. Originally developed for Earth observation (e.g., CYGNSS for hurricane monitoring, TechDemoSat-1 for oceanography), GNSS-R has been adapted for cislunar space, addressing critical gaps in lunar surface analysis (Ruf et al., 2016).

### 5.1 Soil Moisture and Ice Detection

Lunar water ice—concentrated in permanently shadowed regions (PSRs) of the lunar poles—is a critical resource for sustainable exploration, as it can be converted into drinking water, oxygen, and rocket fuel. Detecting and quantifying this ice, however, is challenging: traditional methods (e.g., radar, thermal imaging) struggle to distinguish between surface ice, subsurface ice, and dry regolith. GNSS-R overcomes this by measuring changes in the reflection coefficient of GNSS signals—dry regolith has a low reflection coefficient (≈0.1), while ice-rich regolith has a much higher coefficient (≈0.5), due to differences in dielectric properties (Madsen et al., 2021).

Raney et al. (2022) and Madsen et al. (2021) demonstrated this by analyzing reflected GNSS signals from the lunar surface, showing that GNSS-R can detect subsurface ice up to 1 meter deep with a resolution of 100 meters. The technique works by transmitting a GNSS signal from an Earth-based or lunar-orbiting satellite; the signal reflects off the lunar surface, and a receiver (on the same or a different satellite) captures the reflected wave. By comparing the amplitude and phase of the reflected signal to the direct (unreflected) signal, researchers can infer the presence and concentration of ice. This method is particularly valuable for PSRs, where permanent darkness limits optical imaging and thermal sensors. For example, NASA's LRO could be modified to carry a GNSS-R receiver, enabling systematic mapping of ice in PSRs—data that would guide the selection of landing sites for resource extraction missions.

### 5.2 DEM Generation for Landing Site Analysis

Precise topographical data is essential for selecting safe lunar landing sites, especially in rugged regions (e.g., the lunar south pole) where steep slopes, craters, and boulders pose significant risks. GNSS-R complements traditional DEM generation methods (e.g., LiDAR, stereo imaging) by using time-delay measurements of reflected signals to calculate surface elevation. When a GNSS signal reflects off the lunar surface, the time it takes to travel from the transmitter to the surface to the receiver (the "delay") is proportional to the distance between the transmitter, surface point, and receiver. By measuring this delay for thousands of signal paths, researchers can construct a 3D map of the surface—i.e., a DEM (Garrison, 2018).

Garrison (2018) showed that GNSS-R-derived DEMs have a horizontal resolution of 50–100 meters and a vertical accuracy of ±5 meters for lunar maria, and ±10 meters for highlands—sufficient for initial landing site screening. Unlike LiDAR, which requires a dedicated transmitter and receiver, GNSS-R leverages existing GNSS satellites (e.g., GPS, Beidou), reducing mission cost and complexity. For example, China's Queqiao-2 mission could integrate a GNSS-R receiver to generate DEMs of the lunar far side, where limited prior data increases landing risk (CNSA, 2024). GNSS-R DEMs also support long-term monitoring of lunar surface changes (e.g., crater formation from meteor impacts), providing valuable data for mission planning and geological studies.

**5.3 Space Weather Monitoring**

Space weather—including solar flares, coronal mass ejections (CMEs), and solar wind—poses a threat to cislunar missions, as it can damage spacecraft electronics, disrupt communication, and increase radiation exposure for crew. GNSS-R enables space weather monitoring by measuring signal distortions caused by interactions between GNSS signals and solar plasma or the lunar ionosphere. When a GNSS signal passes through solar plasma (e.g., a CME), the plasma's free electrons delay the signal and alter its phase; similarly, the thin lunar ionosphere (created by solar UV radiation) causes small but measurable signal shifts. By analyzing these distortions, researchers can infer the density and velocity of solar plasma, providing early warning of space weather events (Kintner, 2004).

This application parallels Earth-based ionospheric studies using GNSS phase delays (Kintner, 2004), but extends it to cislunar space—where traditional space weather monitors (e.g., Earth-based telescopes) have limited coverage. For example, a GNSS-R receiver on the Lunar Gateway could track signal distortions from solar plasma, providing real-time data to warn crewed missions of incoming CMEs. This monitoring capability is critical for long-duration lunar habitation, as it enables proactive measures (e.g., shielding activation, mission rescheduling) to mitigate space weather risks.

**5.4 GNSS-R Signal Path and Reflection Mechanism (Block Diagram)**

The GNSS-R system for lunar remote sensing consists of three core components: a GNSS transmitter, a lunar surface target, and a receiver. The signal path unfolds as follows:

Transmission: A GNSS satellite (e.g., GPS Block III) transmits a high-frequency (1.575 GHz for GPS L1) radio signal toward the lunar surface. The signal is encoded with timing and positioning data, enabling precise delay measurements.

Reflection: The signal reaches the lunar surface, where a portion of it reflects off the regolith (or ice-rich regions). The reflection's amplitude and phase depend on the surface's dielectric properties (e.g., ice content) and topography (e.g., slope angle).

Reception: A receiver—mounted on a lunar orbiter (e.g., Queqiao-2), the Lunar Gateway, or a surface rover—captures the reflected signal. The receiver also collects the direct GNSS signal (if in line of sight with the transmitter) for comparison.

Data Processing: The receiver processes the reflected and direct signals, calculating time delays (for topography), amplitude ratios (for ice detection), and phase shifts (for space weather monitoring). This data is transmitted to Earth or processed onboard for real-time mission decisions.

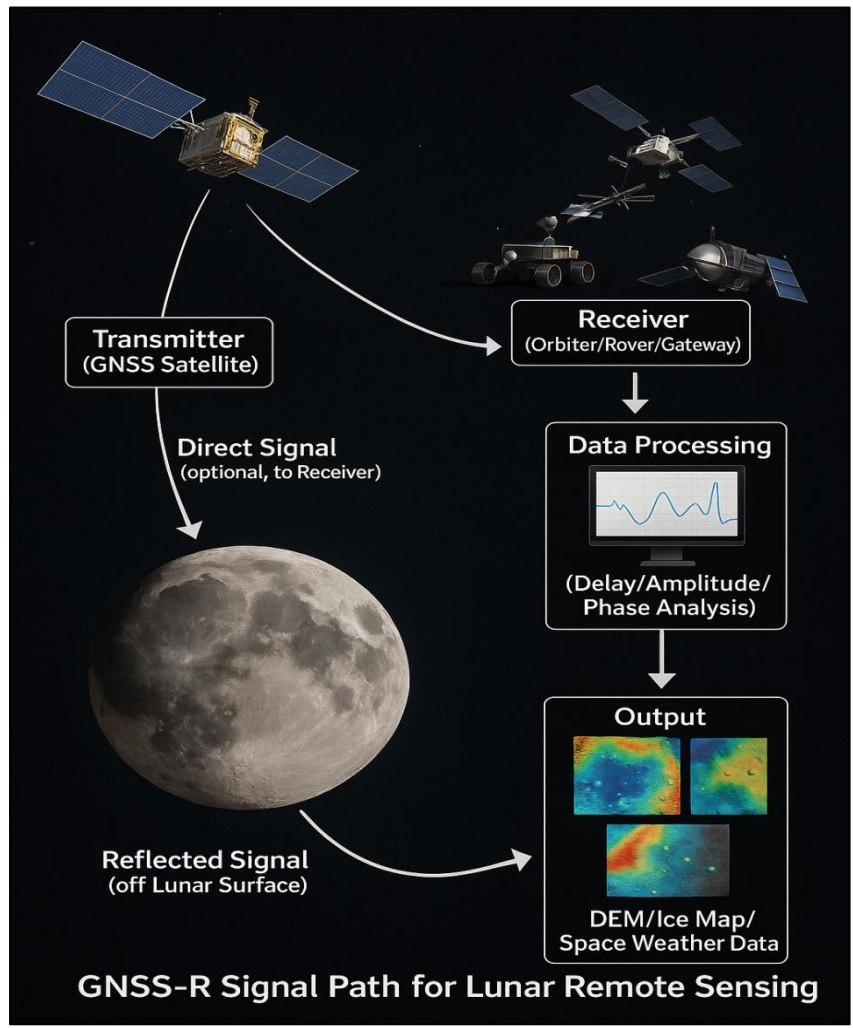

Figure 5: GNSS-R Signal Path for Lunar Remote Sensing

## 6. Challenges and Future Outlook

While the integration of low-energy trajectories, AI/ML, and GNSS-R/PNT has advanced cislunar exploration, significant challenges remain. Addressing these challenges will be critical for realizing sustainable, large-scale cislunar operations. This section outlines key obstacles and future research directions.

### 6.1 Weak GNSS Signal Processing

The 30 dB signal attenuation of Earth-based GNSS beyond GEO remains a primary technical barrier. Even with high-sensitivity receivers, weak signals are prone to noise, interference from solar plasma, and multipath reflections (e.g., off the lunar surface or spacecraft structures), reducing positioning accuracy (Bavaro et al., 2021). Current solutions—such as vector tracking and advanced Kalman filters—improve signal detection but struggle in low-signal-to-noise ratio (SNR) environments. Future research should focus on three areas:

Software-Defined Radios (SDRs): SDRs offer flexible signal processing, enabling adaptive filtering to suppress noise and interference. Integrating SDRs with AI-driven signal detection (e.g., CNNs trained to identify GNSS signals in noise) could improve SNR by 10–15 dB.

Inter-Satellite Link (ISL) Augmentation: Using lunar orbiters as signal relays—equipping them with high-power amplifiers to boost GNSS signals—could reduce attenuation. For example, ESA's Moonlight orbiters could act as "repeaters," amplifying Earth-based GNSS signals before transmitting them to cislunar spacecraft (ESA, 2024).

Multi-Constellation Fusion: Combining signals from multiple GNSS systems (e.g., GPS, Beidou, Galileo) increases the number of available signals, improving geometric coverage and reducing reliance on weak individual signals. Studies by Bavaro et al. (2021) show that multi-constellation fusion can reduce GDOP by 40% beyond GEO.

**6.2 Security, Jamming, and Spoofing in Deep Space**

As cislunar missions become more dependent on GNSS-R/PNT, they become vulnerable to intentional interference—jamming (blocking signals) and spoofing (transmitting fake signals). Unlike near-Earth space, where terrestrial and satellite-based anti-jamming systems exist, cislunar space lacks robust security measures. Jamming from hostile spacecraft or lunar surface assets could disrupt navigation, while spoofing could lead to catastrophic errors (e.g., incorrect lunar orbit insertion). Addressing this requires:

i. Encryption and Authentication: Implementing end-to-end encryption for PNT signals—similar to the GPS Military Code (P(Y) code)—to prevent spoofing. Lunar PNT systems (e.g., LunaNet) should include secure key management protocols to authenticate signals (Garcia et al., 2022).

ii. Redundant Architectures: Designing PNT systems with multiple, independent signal sources (e.g., GNSS, star trackers, IMUs) so that jamming one source does not disable navigation. For example, the Lunar Gateway could use star trackers as a backup to GNSS, ensuring positioning continuity during jamming events.

iii. Interference Detection: Developing AI-driven algorithms to detect jamming/spoofing in real time. These algorithms would analyze signal characteristics (e.g., frequency, phase, amplitude) to identify anomalies, triggering backup systems or evasive maneuvers. Wesson et al. (2021) demonstrated that such algorithms can detect spoofing within 1–2 seconds, minimizing mission risk.

## 6.3 Need for Integrated AI-PNT Systems for Next-Gen Missions

Future cislunar missions—including crewed lunar bases, large-scale resource extraction, and interplanetary staging—will require seamless integration of AI and PNT systems to handle complexity and autonomy. Current systems operate in silos: AI optimizes trajectories, while PNT provides positioning data, with limited real-time communication between them. An integrated AI-PNT system would:

i. Adaptively Optimize Trajectories: Use real-time PNT data (e.g., position, velocity, space weather) to adjust trajectories via DRL, ensuring efficiency and safety. For example, a crewed mission to the lunar south pole could use AI-PNT to avoid unexpected solar flares by rerouting to a sheltered orbit, using PNT data to compute the new path (Scorsoglio et al., 2022).

Ii. Enhance Remote Sensing: Combine AI-driven analysis of GNSS-R data with PNT positioning to create high-fidelity, real-time maps of the lunar surface. For instance, a rover equipped with AI-PNT could use GNSS-R to detect ice deposits and PNT to record their coordinates, generating a dynamic ice map for resource management (Madsen et al., 2021).

Iii. Enable Swarm Operations: Coordinate large satellite swarms (e.g., for global lunar communication) using AI-PNT, which would synchronize orbits, prevent collisions, and optimize coverage. Kolodziejczyk et al. (2020) showed that AI-PNT can maintain sub-meter synchronization for swarms of 50+ satellites, a capability critical for future lunar communication constellations.

To realize integrated AI-PNT systems, researchers must address interoperability (e.g., standardizing data formats between AI and PNT systems) and hardware constraints (e.g., reducing the size/weight/power of onboard AI processors). Additionally, ground-based testing—using lunar simulators and digital twins—will be essential to validate system performance before deployment.

## 7. Conclusion

This review has systematically explored the integration of low-energy Earth-Moon trajectories, AI/ML techniques, and GNSS-R/PNT systems as a unified framework for optimizing cislunar exploration. By analyzing four core transfer paradigms, we highlighted that low-energy trajectories (e.g., WSB-based paths) and spiral transfers offer unparalleled fuel efficiency—reducing propellant costs by up to 50%—while Phase-looping/Pseudo-State models enhance reliability for critical missions such as crewed landings (Belbruno & Miller, 1993; Wilson, 1970; Yue et al., 2009). These trajectory advancements are complemented by AI/ML: CNNs enable rapid, high-precision crater detection and DTM generation, while deep reinforcement learning optimizes real-time trajectory adjustments, increasing mission autonomy and reducing risk (Silburt et al., 2019; Downes et al., 2020; Chen et al., 2022).

The review also demonstrated that GNSS-R/PNT systems overcome the limitations of Earth-based GNSS beyond GEO: augmented architectures (LunaNet, Queqiao-2, Moonlight) provide precise positioning (<10 m for orbiters), while GNSS-R acts as a versatile remote sensing tool—detecting lunar ice, generating DEMs, and monitoring space weather (Garcia et al., 2022; CNSA, 2024; ESA, 2024; Madsen et al., 2021; Garrison, 2018). Together, these technologies address the core challenges of cislunar exploration: cost inefficiency, limited navigation coverage, and inadequate environmental sensing.

**Key insights from this review include:**

Synergy is critical: No single technology solves cislunar challenges—integrating low-energy trajectories (fuel efficiency) with AI (autonomy) and GNSS-R/PNT (navigation/sensing) creates a system that is greater than the sum of its parts.

Autonomy drives sustainability: Onboard AI-PNT systems reduce reliance on the DSN, enabling scalable operations and long-term lunar habitation (Leveque et al., 2021).

GNSS-R is a multi-purpose tool: Beyond navigation, GNSS-R fills critical gaps in lunar environmental sensing, from ice detection to space weather monitoring (Raney et al., 2022; Kintner, 2004).

Future research should prioritize three areas: advancing weak GNSS signal processing (via SDRs and multi-constellation fusion), developing secure anti-jamming/spoofing measures for PNT systems, and integrating AI-PNT hardware to enable seamless autonomy (Bavaro et al., 2021; Wesson et al., 2021; Kolodziejczyk et al., 2020). Additionally, hardware validation—through lunar orbit demonstrations (e.g., modifying the LRO to include a GNSS-R receiver) and ground-based simulators—will be essential to translate theoretical advancements into practical mission capabilities.

As cislunar space evolves into a hub for scientific research, commercial activity, and deep-space staging, the framework presented in this review provides a roadmap for sustainable, cost-effective, and autonomous exploration. By continuing to refine these technologies and foster international collaboration (e.g., standardizing LunaNet-Moonlight interoperability), the global space community can unlock the full potential of cislunar space, paving the way for human missions to Mars and beyond.